\documentclass[sigconf]{acmart}

\usepackage{url}
\usepackage{enumitem}
\usepackage{mwe}
\usepackage{multirow}
\usepackage{multicol}
\usepackage{array}
\newcolumntype{P}[1]{>{\centering\arraybackslash}p{#1}}
\newcolumntype{M}[1]{>{\centering\arraybackslash}m{#1}}
\usepackage{changepage}
\usepackage{subcaption}
\usepackage{cleveref}
\usepackage{tcolorbox}

\usepackage{listings}
\emergencystretch=\maxdimen
\AtBeginDocument{%
  \providecommand\BibTeX{{%
    \normalfont B\kern-0.5em{\scshape i\kern-0.25em b}\kern-0.8em\TeX}}}

\setcopyright{acmcopyright}
\copyrightyear{2023} 
\acmYear{2023} 
\setcopyright{acmlicensed}\acmConference[ITiCSE 2023]{Proceedings of the 2023 Conference on Innovation and Technology in Computer Science Education V. 1}{July 8--12, 2023}{Turku, Finland}
\acmBooktitle{Proceedings of the 2023 Conference on Innovation and Technology in Computer Science Education V. 1 (ITiCSE 2023), July 8--12, 2023, Turku, Finland}
\acmPrice{15.00}
\acmDOI{10.1145/3587102.3588794}
\acmISBN{979-8-4007-0138-2/23/07}

\acmConference[ITICSE '2023]{Conference on Innovation and Technology in Computer Science Education}{July 10--12, 2023}{Turku, Finland}
\begin{document}

\title[ChatGPT, Can You Generate Solutions for my Coding Exercises? ] {ChatGPT, Can You Generate Solutions for my Coding Exercises? An Evaluation on its Effectiveness in an undergraduate Java Programming Course.} 

\author{Eng Lieh Ouh}
\email{elouh@smu.edu.sg}
\affiliation{%
  \institution{Singapore Management University}
  \country{Singapore}
}

\author{Benjamin Kok Siew Gan}
\email{benjamingan@smu.edu.sg}
\affiliation{%
  \institution{Singapore Management University}
  \country{Singapore}
}

\author{Kyong Jin Shim}
\email{kjshim@@smu.edu.sg}
\affiliation{%
  \institution{Singapore Management University}
  \country{Singapore}
}

\author{Swavek Wlodkowski}
\email{swlodkowski@smu.edu.sg}
\affiliation{%
  \institution{Singapore Management University}
  \country{Singapore}
}


\begin{abstract}
In this study, we assess the efficacy of employing the ChatGPT language model to generate solutions for coding exercises within an undergraduate Java programming course. ChatGPT, a large-scale, deep learning-driven natural language processing model, is capable of producing programming code based on textual input. Our evaluation involves analyzing ChatGPT-generated solutions for 80 diverse programming exercises and comparing them to the correct solutions.
Our findings indicate that ChatGPT accurately generates Java programming solutions, which are characterized by high readability and well-structured organization. Additionally, the model can produce alternative, memory-efficient solutions. However, as a natural language processing model, ChatGPT struggles with coding exercises containing non-textual descriptions or class files, leading to invalid solutions.
In conclusion, ChatGPT holds potential as a valuable tool for students seeking to overcome programming challenges and explore alternative approaches to solving coding problems. By understanding its limitations, educators can design coding exercises that minimize the potential for misuse as a cheating aid while maintaining their validity as assessment tools.

\end{abstract}





\maketitle
\vspace{-1em}

\section{Introduction}
ChatGPT, a large-scale deep learning-based natural language processing model, is developed by OpenAI \cite{ChatGPT}. Built on the transformer architecture, this neural network excels in natural language tasks such as language translation, text summarization, question answering, and even programming exercises. Trained on a diverse array of text data, including books, articles, and websites, ChatGPT can generate human-like text and respond effectively to contextual prompts. This makes it an valuable tool for natural language processing tasks like chatbots and virtual assistants, with the potential to revolutionize technology \cite{haque2022think}.

AI-powered language models like ChatGPT are transforming programming education by offering real-time assistance and personalized feedback, enabling students to learn at their own pace and in a manner tailored to their individual needs. However, using such models in assessments raises concerns about examination integrity \cite{susnjak2022chatgpt}. When students access the model during an exam, they could exploit it to generate answers for questions they cannot solve, leading to cheating and invalidating test results.

Programming and object-oriented concepts serve as the cornerstone of computing programs. Numerous research studies have investigated techniques and teaching styles to improve student learning outcomes in programming. These include adopting flipped classrooms \cite{sprint2020improving, mohamed2020evaluating, designingflipped}, enhancing interactivity in video-based programming tutorials \cite{bao2018vt, 10.1145/3510456.3514142, kim2017pedagogical}, and  how teachers can support students in refining their code \cite{keuning2019teachers} to guide novice programmers towards better coding practices. In the era of generative artificial intelligence (AI), Susnjak \cite{susnjak2022chatgpt} investigates concerns regarding online examination integrity with ChatGPT. Susnjak's conclusion emphasizes the importance of educators and institutions acknowledging the potential for ChatGPT-assisted cheating and exploring measures to address it, thus maintaining the fairness and validity of online assessments. 

This study seeks to address the research question (RQ):
\emph{"How effective is ChatGPT in generating coding solutions for Java programming exercises?"}

We evaluate the extent to which ChatGPT-generated solutions meet the requirements of coding exercises in our introductory Java programming course for undergraduate computing students. Each week, students receive programming exercises accompanied by expected outcomes and sample output. During the evaluation, we input the exercise instructions into ChatGPT (versions 3.5 and 4, the latest versions at the time of writing) to generate a solution. The generated programming codes are then executed on our local machine, and the results are compared to the expected outcomes for each exercise.

In Section 2, we provide a literature review focusing on student engagement in programming exercises, followed by an outline of our course design in Section 3. Section 4 offers a detailed account of the weekly exercises and their respective analyses. In Section 5, we summarize the effectiveness of ChatGPT-generated solutions. Finally, Section 6 presents the conclusion of our work.
\section{Background and Literature Review}
Generating code from a high level description of what they should do is a long-standing task in computer science \cite{manna1971toward}. Most prior work has been limited to either restricted domain specific programming languages or short code snippets. In contrast, generating entire programs often relies on understanding the task, reasoning out the appropriate algorithm to solve it, and then writing the code to implement that algorithm. AlphaCode is a system applied to code generation for competitive programming that can generate novel solutions to unseen programming problems \cite{castelvecchi2022chatgpt}. Solving competitive programming problems represents a big step forward. The submission is correct only if it has the correct output on all hidden tests, otherwise it is incorrect. Hidden tests are not visible to the submitter, who must instead write their own tests or rely on the trivial example tests for debugging. Recent AlphaCode researcher analyses indicate that their model was indeed able to solve problems it has never seen before (learning) even though those problems require considerable reasoning.
 
Both ChatGPT and AlphaCode are ‘large language models’ systems based on neural networks that learn to perform a task by digesting massive amounts of existing human-generated text \cite{castelvecchi2022chatgpt}. ChatGPT and AlphaCode are trained on different data sets. While ChatGPT is a general-purpose conversation engine, AlphaCode is specialised and trained exclusively on CodeContest datasets \cite{li2022competition, deepmindcodecontests, alphacodematerials}. Another code generation system is GitHub's Copilot \cite{githubcopilot} which uses the OpenAI Codex. ChatGPT and Codex are OpenAI's products, suggesting that ChatGPT has access to and trained on GitHub public code datasets.
 
Recent popularity of ChatGPT has brought much attention to its benefits (AI pair programming) or problem (cheating) in education. Teachers may have to adjust their assessments by testing in class where they can be monitored or writing questions that required deeper understanding. For example, at Sacred Heart University in Connecticut, teachers will have to give assessments that judge analytical reasoning and not just facts that can be looked up \cite{teachersonalert}. While much more advanced, teachers comparing ChatGPT to programming is similar to the introduction of calculator to mathematics. Besides adjusting the assessments, teachers can educate students about the consequences of cheating during proctored exams and provide students with support so that students do not see the need to cheat \cite{teachersonalert}. In this research, we explore to what extent ChatGPT can be helpful in solving Java programming exercises. Findings from this research can help educators understand how ChatGPT as a "calculator" can 1) enhance Java programming learning for students and 2) craft Java coding exercises so that cheating is minimised.

\section{Course Design and Conduct}
This course is designed for undergraduate students enrolled in computer science and information systems programs. Students are provided with video links to watch and quizzes to self-assess their understanding before attending the class. During class sessions, the instructor conducts coding demonstrations to emphasize key concepts and complement the pre-class materials. Subsequently, students are assigned in-class programming exercises to complete asynchronously. These programming exercises adhere to the requirements of the Oracle Java Foundations Certification \cite{javacert} and are organized into the following learning categories.
\begin{enumerate}[
  align=left,
  leftmargin=2em,
  itemindent=2pt,
  labelsep=0pt,
  labelwidth=2em
]
    \item Java Basics (28 exercises) - Test knowledge on variables, data types, conditionals, loops and collections.
    \item Object-Oriented Concepts (11 exercises) - Test knowledge on object-oriented programming and understanding of API documentation.
    \item Java Classes and Methods (17 exercises) - Test knowledge on method parameters, overloading, inheritance and polymorphism. Understanding of class and sequence diagrams using Unified Modelling Language (UML).
    \item Java Exceptions and I/O (16 exercises) - Test knowledge on how to incorporate exception mechanisms when coding and the use of text files as a persistent storage.
    \item Java Packaging (8 exercises) - Test knowledge on environment variables such as classpath, organise classes into packages and how to use the appropriate modifier for access control.
\end{enumerate}
\section{Evaluation of Java programming exercises}
For each exercise, we copy the exercise instructions and execute them on the ChatGPT \cite{ChatGPT} online. We subscribed to ChatGPT Plus to use both ChatGPT 3.5 and 4 models for this evaluation. We explicitly modify the instructions to include the keyword Java so that ChatGPT does not output codes in other languages. For example, we change "create a method" to "create a Java method". The exercise instructions are in text format. Some questions are supplemented with UML diagrams, compiled class files and/or API documentation to answer the questions. The generated solution is compared with our expected results to evaluate the effectiveness of ChatGPT in answering these exercises. We present the results for each exercise category in the following sections to address our research question.

\subsection{Java Basics}
These exercises require students to declare variables and apply conditionals and loops to implement logic in the main method of the Java source file. We provided expected outputs based on sample input data for students' understanding. Below is an example of the exercise, and Listing \ref{JavaBasics1} shows the two generated solutions from ChatGPT. Both answers fulfilled the exercise requirements.
\\
\begin{adjustwidth}{0.5em}{0pt}
\emph{Exercise 1 - Implement the method sumOfDigits that returns the sum of all the digits of that number. For example, sumOfDigits(123) returns 6 (1 + 2 + 3). For negative numbers, the sum is the same value as that for a positive number. For example, sumOfDigits(-354) returns 12 (3 + 5 + 4).}\\
\end{adjustwidth}
\vspace{-1em}
\noindent
\begin{lstlisting}[caption={Java Basics -  Answers to Exercise 1},captionpos=b,label={JavaBasics1},language=Java, frame=single]
 // Correct Answer 1
public static int sumOfDigits(int n) {
    int sum = 0;
    n = Math.abs(n); // make sure n is positive
    while (n > 0) {
        int digit = n % 10;
        sum += digit;
        n /= 10;
    }
    return sum;
}
 // Correct Answer 2
public static int sumOfDigits(int n) { 
    int sum = 0;
    n = Math.abs(n);
    String numString = String.valueOf(n);
    for (int i = 0; i < numString.length(); i++) {
        sum += Character.getNumericValue(numString.charAt(i));
    }
    return sum;
}
\end{lstlisting}
On exercises to write efficient code, Listing \ref{JavaBasics2} shows the correct output from ChatGPT using StringBuilder. We realise that there is no mention of StringBuilder or comparison between StringBuffer,  + operator and StringBuilder append method. However, this shortfall can be fixed by asking ChatGPT a follow-up question to compare these 3 methods.
\\
\begin{adjustwidth}{0.5em}{0pt}
\emph{Exercise 2 - Write a memory efficient Java Method to concatenate two Strings.}
\end{adjustwidth}
\vspace{-0.5em}
\noindent
\begin{lstlisting}[caption={Java Basics - Answer to Exercise 2},captionpos=b,label={JavaBasics2},language=Java, frame=single]
 // Memory efficient answer 
public static String concatenate(String str1, String str2) {
    StringBuilder sb = new StringBuilder();
    sb.append(str1);
    sb.append(str2);
    return sb.toString();
}
\end{lstlisting}

ChatGPT is capable of generating code that produces the expected output. The following example illustrates an exercise with its expected output, while Listing \ref{JavaBasics3} presents a correct solution generated by ChatGPT.
\\
\begin{adjustwidth}{0.5em}{0pt}
\emph{Exercise 3 - Write Java code that contains only 1 “for” loop to produce the following outputs.
1
12
123
}
\end{adjustwidth}
\noindent
\begin{lstlisting}[caption={Java Bascis - Answer to Exercise 3},captionpos=b,label={JavaBasics3},language=Java, frame=single]
 // Correct Answer
for (int i = 1; i <= 3; i++) {
    for (int j = 1; j <= i; j++) {
        System.out.print(j);
    }
    System.out.println();
}
\end{lstlisting}

For exercises that required more complex output, such as longer instructions to provide a context of the exercise, ChatGPT-3.5 encountered  with incorrect answer but ChatGPT-4 is able to understand the exercise instructions. Below is an example of an exercise and Listing \ref{JavaBasics4} shows an correct output from ChatGPT-4.
\\
\begin{adjustwidth}{0.5em}{0pt}
\emph{Exercise 4 - You need to implement a static method that attempts to align two strings. This method takes in two parameters which are both of type String. Let’s call the first parameter str1 and the second str2. You can assume that str1 is longer than or of the same length as str2. The method should try to align str2 with str1 and insert spaces for additional characters in str1. In the end the method returns true if there is an alignment, and meanwhile the method prints out the alignment. If no alignment can be found, the method returns false and prints out nothing. 
For example, if str1 is "Recess Week" and str2 is "RcWk", then we can align str2 to str1 as follows:\\
Recess Week\\
R\, c\,\,\,\,\,\,      W \,\,   k}\\
\end{adjustwidth}
\vspace{-1em}
\noindent
\begin{lstlisting}[caption={Java Basics - Answer to Exercise 4},captionpos=b,label={JavaBasics4},language=Java, frame=single]
// Correct Answer 
public static boolean matchStrings(String str1, String str2) {
    int str1Len = str1.length();
    int str2Len = str2.length();
    int str2Index = 0;
    StringBuilder alignedStr2 = new StringBuilder();
    for (int str1Index = 0; str1Index < str1Len; str1Index++) {
        if (str2Index < str2Len && str1.charAt(str1Index) == str2.charAt(str2Index)){
            alignedStr2.append(str2.charAt(str2Index));
            str2Index++;
        } else {
            alignedStr2.append(' ');
        }
    }
    if (str2Index == str2Len) { return true;}
    return false;
}
\end{lstlisting}

There can be inconsistencies in the interpretation of program codes in ChatGPT. In one exercise to test the logical operators (\&, \&\&, ||, |) and increment/decrement operators (++). While ChatGPT-3.5 can generate solutions correctly for each type of operation, ChatGPT cannot generate a correct solution with both operations included. Below Listing \ref{JavaBasics41} shows a simplified version of this example. ChatGPT-3.5 is able to learn once we indicate to ChatGPT. However, when we re-evaluate the exercise in a new chat session and modify the conditions, the inconsistency appears again. We understand that the learning requires time to be reflected in new chats. ChatGPT-4 gives the correct answers for this exercise.

\begin{adjustwidth}{0.5em}{0pt}
\emph{Exercise - 5 This exercise explores the differences between short-circuit and normal logical operators. Given the following:}
\end{adjustwidth}
\noindent
\begin{lstlisting}[caption={Java Basics - Operators. Answers to Exercise 5},captionpos=b,label={JavaBasics41},language=Java, frame=single]
// Correct Answer - age is 10
int age = 9;
System.out.println(++age);
// Incorrect Answer for ChatGPT-3.5
// age is 11 but ChatGPT output 10 initially. 
// After repeating the evaluations, 
// ChatGPT acknowledge the output should be 11.
char c = 'a';
int age = 9;
if (c == 'a' & ++age == 10){
    age++;
}
System.out.println(age);
\end{lstlisting}
\vspace{-1em}

\subsection{Object-Oriented Concepts}
These exercises require students to implement classes and methods based on the provided API documentation and, in some cases, class files. Since ChatGPT's input is text-based, we can only submit the exercise instructions. Consequently, ChatGPT is unable to generate fully correct solutions without comprehending the API documentation and class files. Nevertheless, ChatGPT can still produce partial solutions, allowing students a starting point to tackle the problem. Exercise 6 and Listing \ref{OO1} demonstrate a nearly correct solution generated by ChatGPT. However, ChatGPT's performance declines when presented with exercises containing limited text-based information. Exercise 7 and Listing \ref{OO2} display the incorrect outputs generated by ChatGPT alongside the expected solution we anticipate.
\\
\begin{adjustwidth}{0.5em}{0pt}
\emph{Exercise 6 - Study the Shirt API documentation. Write a program called ShirtTest to do the following:\\
a.	Create the following objects using the appropriate constructor: \\
Variable Name	\,\,Brand\,\,\,\,\,\,\,\,\,\,Price	\,\,\,\,Color \\
shirt1	\,\,\,\,\,\,\,\,\,\,\,\,\,\,\,\,\,\,\,\,\,SMUgger	\,\,\,\,85.50\,\,\,\,\,\,Red \\
shirt2	\,\,\,\,\,\,\,\,\,\,\,\,\,\,\,\,\,\,\,\,\,SMUgger	\,\,\,\,90.40\,\,\,\,\,\,Blue \\
shirt3	\,\,\,\,\,\,\,\,\,\,\,\,\,\,\,\,\,\,\,\,\,(no brand) \,\,77.60\,\,\,\,Green \\
b.	Checks whether shirt1 and shirt2 are of the same brand using an appropriate method. \\
c.	Check whether shirt2 and shirt3 are of the same brand using an appropriate method. \\
d.	Set the color of shirt2 to red. \\
e.	Print to console the text representation of shirt1, shirt2 and shirt3.
}
\\
\end{adjustwidth}
\noindent
\begin{lstlisting}[caption={Object Oriented - Answer to Exercise 6},captionpos=b,label={OO1},language=Java , frame=single]
 // Partially Correct Answer 
public class ShirtTest {
    public static void main(String[] args) {
        // a. Create objects 
        Shirt shirt1 = new Shirt("SMUgger", 85.50, "Red");
        Shirt shirt2 = new Shirt("SMUgger", 90.40, "Blue");
        Shirt shirt3 = new Shirt(null, 77.60);
        shirt3.setColor("Green");
        
        // b. Answer is incorrect
        // Method isSameBrand and not sameBrand
        if (shirt1.sameBrand(shirt2)) {
            System.out.println("Shirts 1 and 2 are of the same brand");
        } else {
            System.out.println("Shirts 1 and 2 are not of the same brand");
        }
        
        // c. Same as b. Answer is incorrect
        // Method isSameBrand and not sameBrand
        // Codes left out due to limited pages.
        
        // d. Set the color of shirt2 to red
        shirt2.setColor("Red");
        
        // e. Print the textual representation of shirt1, shirt2, and shirt3
        System.out.println("Shirt 1: " + shirt1);
        System.out.println("Shirt 2: " + shirt2);
        System.out.println("Shirt 3: " + shirt3);
    }
}

\end{lstlisting}

\begin{adjustwidth}{0.5em}{0pt}
\emph{Exercise - 7 You are given the Substance API documentation and class file.  Complete the code for the calculateMass method inside the given file called RadioActiveTest.java}
\end{adjustwidth}
\noindent
\begin{lstlisting}[caption={Object Oriented - Answers to Exercise 7},captionpos=b,label={OO2},language=Java, frame=single]
public class RadioActiveTest {
    // InCorrect Answer from ChatGPT
    public static double calculateMass(double atoms, double atomicWeight) {
        return atoms * atomicWeight;
    }
   // Some codes removed due to limited space.
    // A possible correct answer
    public static double calculateMass(double initialMass, double rateOfDecay, int years) {
        Substance sub = new Substance(initialMass, rateOfDecay);
        for (int i = 0; i < years; i++) {
            sub.decayOneYear();
        }
        return sub.getMass();
    }
}
\end{lstlisting}
\vspace{-1.5em}
\subsection{Java Classes and Methods}
These exercises require students to interpret Unified Modelling Language (UML) class and sequence diagrams to implement the required classes and methods. These exercises challenge ChatGPT with the instructions in the text but the diagrams as pictures. Below is an example of the exercise and Listing \ref{CM1} shows the partially correct output from ChatGPT with missing attributes and methods, as shown in the class diagram.
\\
\begin{adjustwidth}{0.5em}{0pt}
\emph{Exercise 8 - Most men who have completed their National Service with the military know that certain vocations (e.g. commandos) will be given an additional vocation allowance” as a reward for their higher risk. Study the class diagram below. Code the Soldier and Commando classes. The getGrossAllowance() methods should calculate salary using this formula: \\
Soldier’s gross allowance = baseAllowance\\
Commando’s gross allowance = baseAllowance + vocationAllowance
}
\end{adjustwidth}

\includegraphics[scale=0.72]{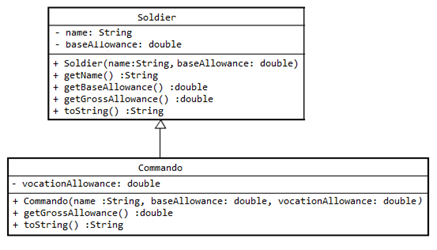}

\noindent
\begin{lstlisting}[caption={Classes and Methods - Answers to Exercise 8},captionpos=b,label={CM1},language=Java, frame=single]
 // Partially Correct Answer from ChatGPT
class Soldier {
    private double baseAllowance;
    public Soldier(double baseAllowance) {
        this.baseAllowance = baseAllowance;
    }
    public double getGrossAllowance() {
        return baseAllowance;
    }
}

class Commando extends Soldier {
    private double vocationAllowance;
    public Commando(double baseAllowance, double vocationAllowance) {
        super(baseAllowance);
        this.vocationAllowance = vocationAllowance;
    }
    public double getGrossAllowance() {
        return super.getGrossAllowance() + vocationAllowance;
    }
}

\end{lstlisting}
\vspace{-1em}
\subsection{Java Packaging}
These exercises evaluate students' comprehension of packaging classes, organizing packaged classes into the proper folders, and defining the javac compile and java execution commands, which include classpath and output folders. When class packages are presented in a class diagram, ChatGPT is unable to determine the package for the class. In some exercises, the Java source file is provided, and we input the source code into ChatGPT. Despite this, the packaging is not accurately recognized. An example of this exercise is provided below.
\\
\begin{adjustwidth}{0.5em}{0pt}
\emph{"Exercise 9  - Place the AnimalFarmTest.java in the proper sub-directories in the sourceFiles folder."}\\
\end{adjustwidth}

\includegraphics[scale=0.44]{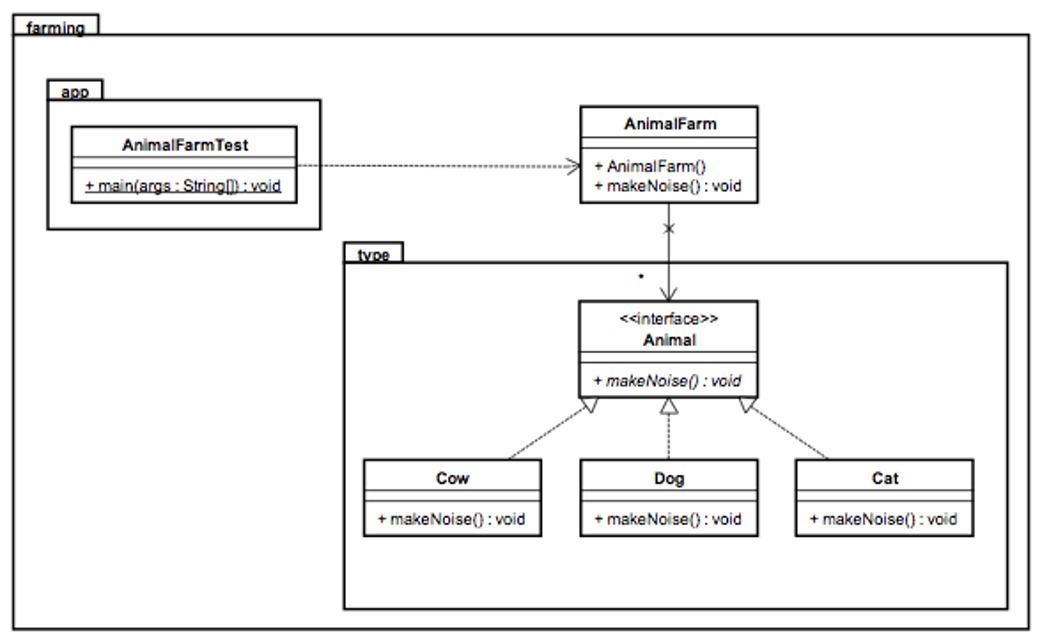}
\vspace{-1em}

\noindent
\begin{lstlisting}[caption={Packaging Sample},captionpos=b,label={package1},language=Java, frame=single]
package farming.app;
import farming.*;

public class AnimalFarmTest {
    public static void main(String[] args) {
        AnimalFarm farm = new AnimalFarm();
        farm.makeNoise();
    }
}
\end{lstlisting}

With both the exercise 9 instructions and Listing \ref{package1}, ChatGPT recommends AnimalFarmTest.java to be placed in sourceFiles/farming/AnimalFarmTest.java instead of sourceFiles/farming/app/AnimalFarmTest.java

\subsection{Java Exceptions and I/O}
These exercises require the student to implement exception classes and codes to handle exceptions. If the exception classes are shown in a class diagram, as shown below, ChatGPT is unable to decipher the package for class, similar to earlier. On the other hand, if the exercise contains existing codes that generate exceptions, ChatGPT can regenerate solutions to handle exceptions. Below is an example of the exercise, and Listing \ref{exception1} shows a correct output from ChatGPT.
\\
\\
\\
\\
\\
\emph{Exercise 10 - The program crashes when an invalid input (e.g. "abc") is entered. Include exception handling into the program to make the program robust, i.e. inform the user of invalid inputs and prompt the user to try again.
}\\
\vspace{-1em}

\noindent
\begin{lstlisting}[caption={Exception - Answer to Exercise 10},captionpos=b,label={exception1},language=Java, frame=single]
 // Correct Answer from ChatGPT
import java.util.InputMismatchException;
import java.util.Scanner;
public class Adder {
    public static void main(String[] args) {
        Scanner sc = new Scanner(System.in);
        boolean valid = false;
        int num1 = 0, num2 = 0;
        while (!valid) {
            try {
                System.out.print("Enter num 1> ");
                num1 = sc.nextInt();

                System.out.println();
                System.out.print("Enter num 2> ");
                num2 = sc.nextInt();
                valid = true;
            } catch (InputMismatchException e) {
                System.out.println("Invalid input. Please enter a valid number.");
                sc.next();
            }
        }
        int sum = num1 + num2;
        System.out.println();
        System.out.println("The answer is " + sum);
    }
}
\end{lstlisting}
\vspace{-1em}
\section{Discussion on the Effectiveness of ChatGPT}
In this section, we summarise the findings from our analysis of ChatGPT's effectiveness in generating coding solutions to address our research question (RQ).
\begin{enumerate}
\item ChatGPT's capability to generate coding solutions positively impacts student learning. Students can obtain guidance on coding solutions, explore alternative methods to solve problems, and compare solutions for efficiency. Educators can instruct students on understanding ChatGPT's features and capabilities from an educational perspective.
\item ChatGPT serves as a valuable resource for students in code debugging. Although assistance from instructors and peers remains helpful, ChatGPT can augment students' self-directed learning abilities by allowing them to independently seek solutions and explore alternative methods. This support is especially beneficial outside of lesson hours or in classroom environments with a high student-to-teacher ratio and also provides timely assistance to the student.
\item ChatGPT-generated solutions may not always be accurate in initial attempts, requiring students to critically evaluate the provided answers. As noted by ChatGPT, it may produce inaccurate information about people, places, or facts. Consequently, students must still rely on their subject knowledge to make their assessments. This evaluation process can contribute to their learning experience. As ChatGPT learns from its mistakes, its performance can improve over time, offering additional benefits to students.
\item As ChatGPT primarily supports text-based inputs, it is less effective for exercises that necessitate the interpretation of API documentation and UML diagrams, such as class and sequence diagrams. Educators can leverage non-text instructions to minimize the likelihood of students copying answers directly from ChatGPT, thus maintaining the integrity of the learning process. We do acknowledge with the advancement in AI technology, non-text inputs can be used. Together with the usage of prompt engineering, these exercises with diagrams can be handled in the future. 
\item  The ChatGPT-generated solutions provide a promising avenue for students to learn asynchronously. We acknowledge this benefit requires students' validation as part of future work.
\end{enumerate}

\section{Conclusion}
The popularity of ChatGPT is growing quickly in recent months. In education, ChatGPT can potentially disrupt how students learn and how educators assess students. With the ChatGPT interactive mode, students can submit their exercise instructions for ChatGPT to answer. In this paper, we evaluate the effectiveness of ChatGPT's ability to generate coding solutions in an undergraduate Java programming course.

Using our course coding exercise instructions as input, ChatGPT generated coding solutions which we evaluate against the expected output. Our evaluation indicates that ChatGPT-generated solutions are effective when the instructions are clear and straightforward. ChatGPT generates partially incorrect solutions when there are complex instructions (e.g. producing a sophisticated representation as output). Due to its support for only text-based inputs, ChatGPT does not yield much benefit to exercises requiring students to interpret API documentation and UML diagrams.

Our analysis shows that ChatGPT can be an effective tool for students to learn asynchronously. They can overcome programming obstacles and learn alternative and more efficient ways to write codes. On the other hand, educators need to be aware of ChatGPT features and develop exercises as well as assessments that minimise students' chance of copying solutions from ChatGPT.

Our future plan involves integrating ChatGPT into our existing video-based programming tutorials \cite{10.1145/3510456.3514142}, allowing students to receive timely guidance on programming challenges such as coding exceptions or logic issues. Additionally, we aim to utilize ChatGPT for summarizing assistance provided by comments in the StackOverflow postings, enabling students to quickly find help on specific topics.
\bibliographystyle{ACM-Reference-Format}
\bibliography{chatgpt}


\begin{thebibliography}{18}


\ifx \showCODEN    \undefined \def \showCODEN     #1{\unskip}     \fi
\ifx \showDOI      \undefined \def \showDOI       #1{#1}\fi
\ifx \showISBNx    \undefined \def \showISBNx     #1{\unskip}     \fi
\ifx \showISBNxiii \undefined \def \showISBNxiii  #1{\unskip}     \fi
\ifx \showISSN     \undefined \def \showISSN      #1{\unskip}     \fi
\ifx \showLCCN     \undefined \def \showLCCN      #1{\unskip}     \fi
\ifx \shownote     \undefined \def \shownote      #1{#1}          \fi
\ifx \showarticletitle \undefined \def \showarticletitle #1{#1}   \fi
\ifx \showURL      \undefined \def \showURL       {\relax}        \fi
\providecommand\bibfield[2]{#2}
\providecommand\bibinfo[2]{#2}
\providecommand\natexlab[1]{#1}
\providecommand\showeprint[2][]{arXiv:#2}

\bibitem[AlphaCode(2023)]%
        {alphacodematerials}
\bibfield{author}{\bibinfo{person}{AlphaCode}.}
  \bibinfo{year}{2023}\natexlab{}.
\newblock \bibinfo{booktitle}{\emph{AlphaCode data materials}}.
\newblock
\urldef\tempurl%
\url{https://zenodo.org/record/6975437#.Y8gBzy8RppQ}
\showURL{%
\tempurl}
\newblock
\shownote{Accessed on 15.01.2023}.


\bibitem[Bao et~al\mbox{.}(2018)]%
        {bao2018vt}
\bibfield{author}{\bibinfo{person}{Lingfeng Bao}, \bibinfo{person}{Zhenchang
  Xing}, \bibinfo{person}{Xin Xia}, {and} \bibinfo{person}{David Lo}.}
  \bibinfo{year}{2018}\natexlab{}.
\newblock \showarticletitle{VT-Revolution: Interactive programming video
  tutorial authoring and watching system}.
\newblock \bibinfo{journal}{\emph{IEEE Transactions on Software Engineering}}
  \bibinfo{volume}{45}, \bibinfo{number}{8} (\bibinfo{year}{2018}),
  \bibinfo{pages}{823--838}.
\newblock


\bibitem[Castelvecchi(2022)]%
        {castelvecchi2022chatgpt}
\bibfield{author}{\bibinfo{person}{Davide Castelvecchi}.}
  \bibinfo{year}{2022}\natexlab{}.
\newblock \showarticletitle{Are ChatGPT and AlphaCode going to replace
  programmers?}
\newblock \bibinfo{journal}{\emph{Nature}} (\bibinfo{year}{2022}).
\newblock


\bibitem[DeepMind(2023)]%
        {deepmindcodecontests}
\bibfield{author}{\bibinfo{person}{DeepMind}.} \bibinfo{year}{2023}\natexlab{}.
\newblock \bibinfo{booktitle}{\emph{DeepMind CodeContests dataset}}.
\newblock
\urldef\tempurl%
\url{https://github.com/deepmind/code_contests}
\showURL{%
\tempurl}
\newblock
\shownote{Accessed on 15.01.2023}.


\bibitem[GitHub(2023)]%
        {githubcopilot}
\bibfield{author}{\bibinfo{person}{GitHub}.} \bibinfo{year}{2023}\natexlab{}.
\newblock \bibinfo{booktitle}{\emph{GitHub Co-Pilot}}.
\newblock
\urldef\tempurl%
\url{https://github.com/features/copilot}
\showURL{%
\tempurl}
\newblock
\shownote{Accessed on 15.01.2023}.


\bibitem[Haque et~al\mbox{.}(2022)]%
        {haque2022think}
\bibfield{author}{\bibinfo{person}{Mubin~Ul Haque}, \bibinfo{person}{Isuru
  Dharmadasa}, \bibinfo{person}{Zarrin~Tasnim Sworna},
  \bibinfo{person}{Roshan~Namal Rajapakse}, {and} \bibinfo{person}{Hussain
  Ahmad}.} \bibinfo{year}{2022}\natexlab{}.
\newblock \showarticletitle{" I think this is the most disruptive technology":
  Exploring Sentiments of ChatGPT Early Adopters using Twitter Data}.
\newblock \bibinfo{journal}{\emph{arXiv preprint arXiv:2212.05856}}
  (\bibinfo{year}{2022}).
\newblock


\bibitem[Keuning et~al\mbox{.}(2019)]%
        {keuning2019teachers}
\bibfield{author}{\bibinfo{person}{Hieke Keuning}, \bibinfo{person}{Bastiaan
  Heeren}, {and} \bibinfo{person}{Johan Jeuring}.}
  \bibinfo{year}{2019}\natexlab{}.
\newblock \showarticletitle{How teachers would help students to improve their
  code}. In \bibinfo{booktitle}{\emph{Proceedings of the 2019 ACM Conference on
  Innovation and Technology in Computer Science Education}}.
  \bibinfo{pages}{119--125}.
\newblock


\bibitem[Kim and Ko(2017)]%
        {kim2017pedagogical}
\bibfield{author}{\bibinfo{person}{Ada~S Kim} {and} \bibinfo{person}{Amy~J
  Ko}.} \bibinfo{year}{2017}\natexlab{}.
\newblock \showarticletitle{A pedagogical analysis of online coding tutorials}.
  In \bibinfo{booktitle}{\emph{Proceedings of the 2017 ACM SIGCSE Technical
  Symposium on Computer Science Education}}. \bibinfo{pages}{321--326}.
\newblock


\bibitem[Li et~al\mbox{.}(2022)]%
        {li2022competition}
\bibfield{author}{\bibinfo{person}{Yujia Li}, \bibinfo{person}{David Choi},
  \bibinfo{person}{Junyoung Chung}, \bibinfo{person}{Nate Kushman},
  \bibinfo{person}{Julian Schrittwieser}, \bibinfo{person}{R{\'e}mi Leblond},
  \bibinfo{person}{Tom Eccles}, \bibinfo{person}{James Keeling},
  \bibinfo{person}{Felix Gimeno}, \bibinfo{person}{Agustin Dal~Lago},
  {et~al\mbox{.}}} \bibinfo{year}{2022}\natexlab{}.
\newblock \showarticletitle{Competition-level code generation with alphacode}.
\newblock \bibinfo{journal}{\emph{Science}} \bibinfo{volume}{378},
  \bibinfo{number}{6624} (\bibinfo{year}{2022}), \bibinfo{pages}{1092--1097}.
\newblock


\bibitem[Manna and Waldinger(1971)]%
        {manna1971toward}
\bibfield{author}{\bibinfo{person}{Zohar Manna} {and}
  \bibinfo{person}{Richard~J Waldinger}.} \bibinfo{year}{1971}\natexlab{}.
\newblock \showarticletitle{Toward automatic program synthesis}.
\newblock \bibinfo{journal}{\emph{Commun. ACM}} \bibinfo{volume}{14},
  \bibinfo{number}{3} (\bibinfo{year}{1971}), \bibinfo{pages}{151--165}.
\newblock


\bibitem[Meckler and Verma(2022)]%
        {teachersonalert}
\bibfield{author}{\bibinfo{person}{Laura Meckler} {and}
  \bibinfo{person}{Pranshu Verma}.} \bibinfo{year}{2022}\natexlab{}.
\newblock \bibinfo{booktitle}{\emph{Teachers are on alert for inevitable
  cheating after release of ChatGPT.}}
\newblock
\urldef\tempurl%
\url{https://www.washingtonpost.com/education/2022/12/28/chatbot-cheating-ai-chatbotgpt-teachers/}
\showURL{%
\tempurl}
\newblock
\shownote{Accessed on 15.01.2023}.


\bibitem[Mohamed(2020)]%
        {mohamed2020evaluating}
\bibfield{author}{\bibinfo{person}{Abdallah Mohamed}.}
  \bibinfo{year}{2020}\natexlab{}.
\newblock \showarticletitle{Evaluating the Effectiveness of Flipped Teaching in
  a Mixed-Ability CS1 Course}. In \bibinfo{booktitle}{\emph{Proceedings of the
  2020 ACM Conference on Innovation and Technology in Computer Science
  Education}} (Trondheim, Norway) \emph{(\bibinfo{series}{ITiCSE '20})}.
  \bibinfo{publisher}{Association for Computing Machinery},
  \bibinfo{address}{New York, NY, USA}, \bibinfo{pages}{452–458}.
\newblock
\showISBNx{9781450368742}
\urldef\tempurl%
\url{https://doi.org/10.1145/3341525.3387395}
\showDOI{\tempurl}


\bibitem[OpenAI(2023)]%
        {ChatGPT}
\bibfield{author}{\bibinfo{person}{OpenAI}.} \bibinfo{year}{2023}\natexlab{}.
\newblock \bibinfo{booktitle}{\emph{{ChatGPT: Optimizing Language Models for
  Dialogue }}}.
\newblock
\urldef\tempurl%
\url{https://openai.com/blog/chatgpt/}
\showURL{%
\tempurl}
\newblock
\shownote{Accessed on 15.01.2023}.


\bibitem[Oracle(2023)]%
        {javacert}
\bibfield{author}{\bibinfo{person}{Oracle}.} \bibinfo{year}{2023}\natexlab{}.
\newblock \bibinfo{booktitle}{\emph{Java Foundations Exam Number: 1Z0-811}}.
\newblock
\urldef\tempurl%
\url{https://education.oracle.com/java-foundations/pexam_1Z0-811}
\showURL{%
\tempurl}
\newblock
\shownote{Accessed on 15.01.2023}.


\bibitem[Ouh and Gan(2022)]%
        {designingflipped}
\bibfield{author}{\bibinfo{person}{Eng~Lieh Ouh} {and}
  \bibinfo{person}{Benjamin Kok~Siew Gan}.} \bibinfo{year}{2022}\natexlab{}.
\newblock \showarticletitle{Designing flipped learning activities for beginner
  programming course.}. In \bibinfo{booktitle}{\emph{Proceedings of 28th AMCIS:
  Innovative Research Informing Practice}}. \bibinfo{address}{Minneapolis,
  USA}.
\newblock


\bibitem[Ouh et~al\mbox{.}(2022)]%
        {10.1145/3510456.3514142}
\bibfield{author}{\bibinfo{person}{Eng~Lieh Ouh}, \bibinfo{person}{Benjamin
  Kok~Siew Gan}, {and} \bibinfo{person}{David Lo}.}
  \bibinfo{year}{2022}\natexlab{}.
\newblock \showarticletitle{ITSS: Interactive Web-Based Authoring and Playback
  Integrated Environment for Programming Tutorials}. In
  \bibinfo{booktitle}{\emph{Proceedings of the ACM/IEEE 44th International
  Conference on Software Engineering: Software Engineering Education and
  Training}} (Pittsburgh, Pennsylvania) \emph{(\bibinfo{series}{ICSE-SEET
  '22})}. \bibinfo{publisher}{Association for Computing Machinery},
  \bibinfo{address}{New York, NY, USA}, \bibinfo{pages}{158–164}.
\newblock
\showISBNx{9781450392259}
\urldef\tempurl%
\url{https://doi.org/10.1145/3510456.3514142}
\showDOI{\tempurl}


\bibitem[Sprint and Fox(2020)]%
        {sprint2020improving}
\bibfield{author}{\bibinfo{person}{Gina Sprint} {and} \bibinfo{person}{Erik
  Fox}.} \bibinfo{year}{2020}\natexlab{}.
\newblock \showarticletitle{Improving student study choices in CS1 with
  gamification and flipped classrooms}. In
  \bibinfo{booktitle}{\emph{Proceedings of the 51st ACM Technical Symposium on
  Computer Science Education}}. \bibinfo{pages}{773--779}.
\newblock


\bibitem[Susnjak(2022)]%
        {susnjak2022chatgpt}
\bibfield{author}{\bibinfo{person}{Teo Susnjak}.}
  \bibinfo{year}{2022}\natexlab{}.
\newblock \showarticletitle{ChatGPT: The End of Online Exam Integrity?}
\newblock \bibinfo{journal}{\emph{arXiv preprint arXiv:2212.09292}}
  (\bibinfo{year}{2022}).
\newblock


\end{thebibliography}


\end{document}